\documentclass[aip,amsmath,amssymb]{revtex4-2}

\usepackage{graphicx}
\usepackage{dcolumn}
\usepackage{bm}

\usepackage[utf8]{inputenc}
\usepackage[T1]{fontenc}
\usepackage{mathptmx}
\usepackage{etoolbox}

\makeatletter
\def\@email#1#2{%
 \endgroup
 \patchcmd{\titleblock@produce}
  {\frontmatter@RRAPformat}
  {\frontmatter@RRAPformat{\produce@RRAP{*#1\href{mailto:#2}{#2}}}\frontmatter@RRAPformat}
  {}{}
}%

\makeatother
\begin{document}

\preprint{AIP}

\title{A Partition Function Estimator}
\author{Ying-Chih Chiang}
 \affiliation{Kobilka Institute of Innovative Drug Discovery, School of Medicine, The Chinese University of Hong Kong, Shenzhen, 2001 Longxiang Boulevard, 518172, Shenzhen, China}
 \email{chiangyc@cuhk.edu.cn}

\author{Frank Otto}%
 \affiliation{Advanced Research Computing Centre, University College London, WC1H 9BT, UK}
 \email{f.otto@ucl.ac.uk}


\author{Jonathan W. Essex}
 \affiliation{Department of Chemistry, University of Southampton, SO17 1BJ, UK}

\date{\today}

\begin{abstract}
We propose a simple estimator that allows to calculate 
the absolute value of a system's partition function
from a finite sampling of its canonical ensemble.
The estimator utilizes a volume correction term to compensate
the effect that the finite sampling cannot cover the whole
configuration space.
As a proof of concept, the estimator is applied to calculate 
the partition function for several model systems, and 
the results are compared with the numerically exact solutions.
Excellent agreement is found, demonstrating that 
a solution for an efficient calculation of partition functions
is possible. 
\end{abstract}

\maketitle

\section{Introduction}
How to calculate the partition function of a system has long been 
an important question in physics because many quantities, such as
the free energy or the entropy, can then be determined afterwards.
In the past two decades, the question has been answered with 
various methods, such as the Wang-Landau algorithm~\cite{WangLandau} and nested sampling~\cite{Skilling2004,Skilling2006,NS2022Review}. 
Both methods compute the (cumulative) density of states of a system
and allow a direct integration over the Boltzmann factor to yield 
the partition function as well as other quantities of interest. 
In particular, the nested sampling algorithm has demonstrated
its power in studying thermodynamic properties in material science~\cite{NS2021Review}. 
The tested systems include Lennard-Jones clusters~\cite{Partay2010},
metallic systems~\cite{Partay2018CMS,Partay2020,Partay2022}, alloys~\cite{Partay2018SE}, 
and small water clusters~\cite{Partay2019}.
In these examples, the partition function is rarely the sole goal 
of the study, but rather a basic property that one can calculate 
apart from the heat capacity or a phase-diagram. 

However, calculating the partition function itself is still meaningful, 
since the ratio between two partition functions indicates 
the free energy difference between two systems (or two states). 
For instance, comparing the partition function of a system with
a protein and a free ligand to the partition function of the protein-ligand complex
yields the binding free energy between the protein and the ligand. 
This information can guide the design of drug molecules during lead optimization.
Yet, directly calculating the partition function is computationally expensive. 
Since the desired property is not a single partition function but the ratio between two, 
calculating this ratio using theories tailored to save computational resources seems more reasonable.
For instance, using importance sampling, the ratio between
two partition functions can be obtained by a single sampling on one
of the states, a method known as Zwanzig's equation~\cite{Zwanzig1954}
or free energy perturbation (FEP)~\cite{Chipot2010}. 
Together with post-processing methods such as Bennett's acceptance ratio (BAR)~\cite{BAR} 
or multistate Bennett acceptance ratio (MBAR)~\cite{MBAR} to combine data 
from sampling over different states, multistep free energy 
perturbation (mFEP)~\cite{mFEP} has became a standard approach for 
binding free energy calculation.
Other commonly used free energy methods include Kirkwood's thermodynamic integration~\cite{Kirkwood1935},
the Jarzynski equality~\cite{Jarzynski1997}, etc. 
A comprehensive review can be found in the literature~\cite{FEP_book}. 
These free energy methods enable e.g.\ the calculation of the binding
free energy between a protein and a ligand~\cite{Karplus2003,Roux2005,Mobley2007,Wang2015},
between two proteins~\cite{Gumbart2013}, or for evaluating the permeability of ligands 
through a lipid bilayer~\cite{Comer2017}.
Today, free energy calculations have become an indispensable tool in computer-aided drug design~\cite{Cournia2017,Muegge2023,York2023,Moore2023}.

While the aforementioned achievements makes one question the need for
calculating a single partition function directly, one may still wonder 
whether it is possible to calculate a single partition function more effectively, 
perhaps reaching a level where the performance of the calculation is comparable 
to these advanced methods, at least in some simple systems. 
To answer this question, we propose a Partition Function Estimator 
(PFE) that can compute the value of a single partition function from a 
finite sampling.  As a proof of concept, we apply this theory 
to several model examples, including the one-dimensional harmonic 
oscillator potential, the one-dimensional double-well potential, 
the two-dimensional M\"uller-Brown potential, and up to 30
Lennard-Jones particles in a three-dimensional box.
These examples are chosen because their references can be easily obtained:
For model potentials, exact numerical solutions are available via
brute-force integration, while the partition functions of 
Lennard-Jones particles can be obtained using nested sampling
or standard methods such as mFEP-MBAR.

\section{Theory and Methods}

\subsection{The Partition Function Estimator}
Consider a system composed of $N$ identical particles of mass $m$. 
The Hamiltonian reads,
\begin{eqnarray}
\label{eq:Hamiltonian}
H(\mathbf{p},\mathbf{q}) = \frac{\mathbf{p}^2}{2m} + U(\mathbf{q}) \,
\end{eqnarray}
where $\mathbf{p}$ and $\mathbf{q}$ denote the momenta and the spatial
coordinates of the particles, respectively. $U(\mathbf{q})$ denotes
the potential of the system.
The canonical partition function $Z$ of the system is defined as,
\begin{eqnarray}
\label{eq:Z}
Z = \frac{1}{N!\,h^{3N}}\int e^{-\beta H(\mathbf{p},\mathbf{q})} \, \text{d} \mathbf{p} \, \text{d} \mathbf{q}  \,,
\end{eqnarray}
where $h$ is Planck's constant, and $\beta$ denotes the 
inverse temperature $T$ multiplied with the Boltzmann constant
$k_B$ ($\beta=1/k_BT$).
$Z$ is a function of $\beta$, but this dependency is
dropped for ease of notation.
The integration over phase space in Eq.~\ref{eq:Z} is separable,
and the integration over the momenta can be performed analytically.
The partition function of the system hence can be rewritten as~\cite{NS2022Review},
\begin{eqnarray}
\label{eq:Z2}
Z = \frac{1}{N!}\left( \frac{2\pi m}{\beta h^2} \right)^{3N/2} 
\int e^{-\beta U(\mathbf{q})} \, \text{d} \mathbf{q} \;.
\end{eqnarray}
With the integration over momentum space carried out, we can focus on 
the integration over coordinate space, and thus define
\begin{eqnarray}
\label{eq:Q}
Q = \int e^{-\beta U(\mathbf{q})} \, \text{d} \mathbf{q} \;,
\end{eqnarray}
to denote the spatial contribution to the partition function $Z$.
In the same vein, the expectation value for any function $f$ that
depends solely on the coordinates $\mathbf{q}$, can be expressed as
\begin{eqnarray*}
    \langle f \rangle = \int f(\mathbf{q}) \frac{e^{-\beta U(\mathbf{q})}}{Q} \text{d} \mathbf{q} \;.
\end{eqnarray*}
Our discussion below will focus on how $Q$ can be calculated from 
a finite sampling of the system's canonical ensemble. 

\begin{figure}
\includegraphics[width=0.5\textwidth]{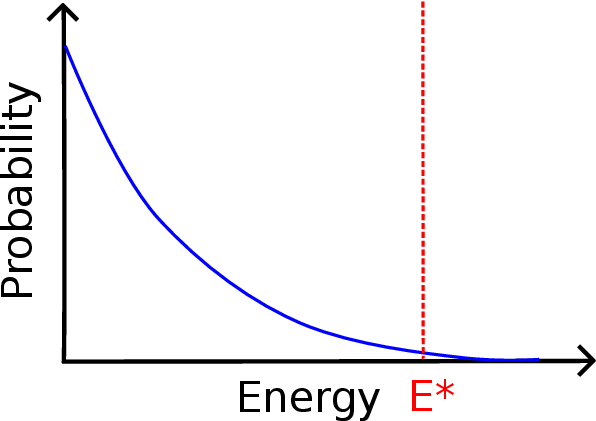}
\caption{\label{fig:illustration}%
A sketch of the energy distribution for the canonical
ensemble of a finite system. Owing to the exponential nature of
the Boltzmann factor, the probability for sampling microstates
with high energy will be very low.  Above a certain energy
$E^{*}$ the sampling is considered to be insufficient.
}
\end{figure}

When sampling a system at finite temperature, the probability for
finding a microstate $i$ with energy $E_i$ is proportional to
$\exp(-\beta E_i)$, and consequently the probability for finding
\emph{any} microstate with energy $E$ is proportional to
$g(E) \exp(-\beta E)$ where $g(E)$ denotes the density of states.
Due to the exponential nature of the Boltzmann factor, the energy
distribution of the system eventually decreases with increasing
energy -- see Fig.~\ref{fig:illustration} for an illustration --
such that above a certain energy level $E^{*}$, the sampling becomes
insufficient, i.e.\ it is impractical to obtain enough samples with
energy $E > E^{*}$ to reproduce the actual probability distribution
in this energy regime.
In practice, this is unproblematic if one is interested in the
sample average of quantities that don't grow exponentially with
$E$, such as the system energy itself and most other quantities
of physical interest. For these, the sample average is insensitive
to the high-energy tail of the probability distribution, and it is
of no concern if it was captured insufficiently during sampling.

Yet, in the extreme case one could be interested in evaluating
the sample average of the inverse Boltzmann factor $b(E) = \exp(+\beta E)$.  Its
theoretical expectation value is given by
\begin{eqnarray}
\label{eq:bad_f}
\langle b(E) \rangle
= \int e^{+\beta U(\mathbf{q})} \frac{e^{-\beta U(\mathbf{q})}}{Q} \text{d} \mathbf{q}
= \frac{L^{3N}}{Q} \;,
\end{eqnarray}
where $L^3$ is the volume of a cubic box that the particles are confined to.
If we could obtain an estimate for $\langle b (E) \rangle$ via sampling, then
Eq.~\ref{eq:bad_f} could be readily used to calculate $Q$. However,
in a sample of size $n$ with energies $E_i$ distributed according to
the distribution $p(E) \propto \exp(-\beta E)$,
the sample average of $b$ is given by
\begin{eqnarray}
\label{eq:b_bar}
\bar{b} = \frac{1}{n} \sum_{i=1}^{n} e^{+\beta E_i} \;,
\end{eqnarray}
which is extremely sensitive to the high-energy samples.
In practice, the insufficient sampling of the high-energy tail leads to a very
large fluctuation of $\bar{b}$, making it useless as an estimate for
$\langle b \rangle$ and thus for determining $Q$.
Note that here as well as below, we clearly distinguish between the
expectation value ($\langle b \rangle$) and the sample average ($\bar{b}$);
while the former is an exact (theoretical) value, the latter is empirically
obtained from a finite sample and will fluctuate if the sampling is repeated,
perhaps even failing to converge in case of insufficient sampling.

Let us now consider another function $f(E;E^{*})$ that reads,
\begin{eqnarray}
\label{eq:function_f}
f(E;E^{*}) = e^{+\beta E} \, \theta(E^{*}-E) \;,
\end{eqnarray}
where $\theta(E^{*}-E)$ denotes the Heaviside step function, 
which is 1 for $E \leq E^{*}$ and 0 for $E > E^{*}$.
In short, $f(E;E^{*})$ is the inverse Boltzmann factor,
but truncated to zero for energies $E$ larger than some
chosen parameter $E^{*}$.
With $E$ set to the potential energy, i.e.\ $E=U(\mathbf{q})$,
the expectation value of $f$ is then given by,
\begin{eqnarray}
\label{eq:expectation}
\langle f(E;E^{*}) \rangle 
= \int e^{+\beta E} \, \theta(E^{*}-E) \frac{e^{-\beta U(\mathbf{q})}}{Q} \, \text{d} \mathbf{q}
= \frac{V(E^{*})}{Q} \;,
\end{eqnarray}
with $V(E^{*})$ defined as,
\begin{eqnarray}
\label{eq:volume}
V(E^{*}) = \int \theta(E^{*} - U(\mathbf{q})) \, \text{d} \mathbf{q} \;,
\end{eqnarray}
which is the volume of the coordinate space where the potential 
energy is smaller than $E^{*}$.
Consequently, $Q$ can be obtained via,
\begin{eqnarray}
\label{eq:pfe}
\ln Q = \ln V(E^{*}) \,-\, \ln \langle f(E;E^{*}) \rangle \;.
\end{eqnarray}
Eq.~\ref{eq:pfe} is the proposed estimator.
We stress that this equation itself is exact, in that no approximations have
been made, and holds for any value of $E^{*}$.
Obviously, to make use of this equation, we still need to determine values
for both $\langle f(E;E^{*}) \rangle$ and $V(E^{*})$, along with
finding a ``good'' value of $E^{*}$.

\subsection{Finding $E^{*}$}

Let us first concentrate on the expectation value $\langle f(E;E^{*}) \rangle$.
We wish to use the sample average of $f$, i.e.
\begin{eqnarray}
\label{eq:f_bar}
\bar{f} 
  = \frac{1}{n} \sum_{i=1}^{n} e^{+\beta E_i} \theta(E^{*} - E_i) \;,
\end{eqnarray}
as an estimate for $\langle f(E;E^{*}) \rangle$.  In contrast to Eq.~\ref{eq:b_bar},
this sample average is not sensitive to the high-energy tail, provided that
$E^{*}$ is chosen such that energies below $E^{*}$ are all sufficiently sampled, 
cf.\ Fig.~\ref{fig:illustration}.
Indeed, if $E^{*}$ were chosen to be vary large, then Eq.~\ref{eq:f_bar}
would suffer from strong fluctuation due to insufficient sampling.
On the other hand, if $E^{*}$ were chosen to be very small, then the
number of samples that contribute effectively to Eq.~\ref{eq:f_bar}
would be very small, which again increases its error.
This suggests that there is an optimal choice for $E^{*}$ that
can minimize the relative error in $\bar{f}$.

According to Eq.~\ref{eq:pfe}, $\ln Q$ is given by the difference between
$\ln V(E^{*})$ and $\ln \langle f(E;E^{*}) \rangle$. We are now using
$\ln \bar{f}$ as an estimate for the second term. Since this is the term
that comes from the sample average (we will discuss $\ln V(E^{*})$ in the next section),
it makes sense to choose an $E^*$ that can minimize the standard deviation of
$\ln \bar{f}$.  This is given by,
\begin{eqnarray}
\label{eq:error}
\sigma_{M} = \sqrt{ \frac{1}{n} \,
  \frac{\langle f(E;E^{*})^2 \rangle - \langle f(E;E^{*}) \rangle^2 }{\langle f(E;E^{*}) \rangle^2}}  \;,
\end{eqnarray}
which is just the relative standard error of $f(E;E^{*})$, with $n$ being the number of samples.
To minimize this error, we find $E^*$ such that
\begin{eqnarray}
\label{eq:variation}
\frac{\partial \sigma_M^2}{\partial E^{*}} =
\frac{\partial}{\partial E^{*}} \left[ \frac{1}{n}
\left( \frac{\langle f(E;E^{*})^2 \rangle}{\langle f(E;E^{*}) \rangle^2}  -1 \right) \right]
= 0  \;.
\end{eqnarray}
Expressing the expectation values in the above equation as integrals over energy space
by utilizing the density of states $g(E)$, we have
\begin{eqnarray}
\label{eq:avg}
\langle f(E;E^{*}) \rangle = \int e^{+\beta E} \theta(E^{*}-E) \, g(E) \frac{e^{-\beta E}}{Q} \, \text{d} E = \frac{1}{Q} \int \theta(E^{*}-E) \, g(E) \, \text{d} E \;,
\end{eqnarray}
and
\begin{eqnarray}
\label{eq:avg2}
\langle f(E;E^{*})^2 \rangle = \int e^{+2\beta E} \theta(E^{*}-E)^2 \, g(E) \frac{e^{-\beta E}}{Q} \, \text{d} E = \frac{1}{Q} \int e^{+\beta E} \theta(E^{*}-E) \, g(E) \, \text{d} E \;.
\end{eqnarray}
Note that in the above equation, we have used the property of the Heaviside
function that $\theta^2 = \theta$. Further, the derivative of the Heaviside
function gives the Dirac delta function,
\begin{eqnarray}
\label{eq:Dirac_delta}
\frac{\partial}{\partial E^{*}} \theta(E^{*}-E) = \delta(E^*{}-E) \;,
\end{eqnarray}
which is only non-zero when $E = E^{*}$.
Using Eqs.~\ref{eq:avg}-\ref{eq:Dirac_delta}, Eq.~\ref{eq:variation} results
in the condition
\begin{eqnarray*}
0 &=& 
\frac{1}{\langle f(E;E^{*}) \rangle^2} \frac{\partial}{\partial E^{*}} \langle f(E;E^{*})^2 \rangle
\,-\, 2\, \frac{\langle f(E;E^{*})^2 \rangle}{\langle f(E;E^{*}) \rangle^3}
  \frac{\partial}{\partial E^{*}} \langle f(E;E^{*}) \rangle
\\
&=& \frac{1}{\langle f(E;E^{*}) \rangle^2} \left[
\frac{g(E^{*})}{Q} e^{+\beta E^{*}} \,-\,
2\, \frac{\langle f(E;E^{*})^2 \rangle}{\langle f(E;E^{*}) \rangle} \frac{g(E^{*})}{Q}
\right] \;,
\end{eqnarray*}
or after simplification,
\begin{eqnarray}
\label{eq:E*exact}
e^{+ \beta E^{*}} = 2\, \frac{\langle f(E;E^{*})^2 \rangle}{ \langle f(E;E^{*}) \rangle} \;.
\end{eqnarray}

According to Eq.~\ref{eq:E*exact}, $E^{*}$ depends on the expectation values of
$f(E;E^{*})$ and $f(E;E^{*})^2$. Both can be estimated from
their respective sample averages. However, as both also depend on
$E^{*}$, this equation must be solved iteratively: 
First, one collects $n$ samples e.g.\ via Monte Carlo sampling,
and saves the trajectory and the associated energies $E_i$.
Then, one calculates $\bar{f}$ via Eq.~\ref{eq:f_bar}
and similarly $\overline{f^2}$, using an initial guess for $E^{*}$,
e.g. the maximum energy encountered during sampling.
Afterwards, $E^{*}$ is updated via
\begin{eqnarray}
\label{eq:E*}
    E^{*} = \frac{1}{\beta} \left( \ln 2 + \ln \overline{f^2} - \ln \bar{f} \right) \;,
\end{eqnarray}
and this process is repeated until $E^{*}$ converges.
Notably, each iteration can employ the same trajectory; one
merely has to update $E^{*}$ and recalculate the sample averages,
which is computationally cheap.
Finally, one can use the optimized $E^{*}$ to estimate
$\langle f(E;E^{*})\rangle$ and $\sigma_M$ from
$\bar{f}$ and $\overline{f^2}$.

\subsection{Calculating $V(E^{*})$}

With $E^{*}$ determined, we can turn to the calculation of the volume term 
$V(E^{*})$. According to Eq.~\ref{eq:volume}, $V(E^{*})$ is the 
volume of the coordinate space with $E < E^{*}$. 
For simple low-dimensional models, such as a harmonic oscillator or 
M\"uller-Brown potential, this volume can be calculated directly 
via ``binning'', i.e.\ dividing the space into small bins,
assigning each sample to its corresponding bin, and simply
counting the number of occupied bins.
However, as the dimensionality increases, no trajectory
could cover the entire admissible coordinate space via binning.
Calculating $V(E^{*})$ by this or other na\"ive integration
approaches is hence not possible.
Second, $V(E^{*})$ soon becomes very small in comparison with the total volume
of the coordinate space. Thus a simple Monte Carlo integration would also be
insufficient to address this issue.

Fortunately, the nested sampling algorithm~\cite{Skilling2004,NS2022Review} was designed 
to tackle such a problem.
Here, we employ a slightly modified version of nested sampling, making it more
suitable for the purpose of Eq.~\ref{eq:volume}.  
Briefly, the implemented algorithm is as follows:
\begin{itemize}
    \item[0.] Randomly generate $N$ ``walkers'' with energies smaller than a starting energy ceiling $E_0$. Each walker is an independent copy of the system that will be propagated in the course of the algorithm. The probability distribution for the initial walkers must be uniform over the coordinate space. The starting energy ceiling in principle should be the highest potential energy possible, but for practical purposes it is normally set to a very large value, such as $10^{12} k_BT$. The initial volume $V_0$ is then given by the volume of the entire coordinate space, $V_0 = L^{3N}$.
    \item[1.] Select a fraction $p$ ($0 < p < 1$). This determines the new energy ceiling for the current ($i$-th) iteration, $E_i = p \cdot E_{i-1}$.  Use a simple Monte Carlo integration to calculate the volume $V(E_i)$, i.e.\ the volume of coordinate space with energy smaller than $E_i$. Assuming that $n_i$ walkers have energies below $E_i$, we have $V(E_i) = (n_i/N) \cdot V(E_{i-1})$.
    \item[2.] Relax the walkers whose energies are larger than $E_i$, and propagate them freely below this energy, until they represent a sample with uniform probability over the coordinate space with energy $E < E_i$. Rejection sampling is employed for this step.
    \item[3.] Repeat Steps 1 and 2 until $E_i < E^{*}$ is reached. As it is unlikely for $E_i$ to exactly hit $E^{*}$, interpolation may be required to determine $V(E^{*})$ from $V(E_i)$ and $V(E_{i-1})$.
\end{itemize}

Compared to the original nested sampling, our implementation is more akin
to an iterative Monte Carlo integration.  Our approach differs 
from nested sampling in two aspects. First and most importantly, 
nested sampling 
throws away a fixed number of walkers in each iteration,
leading to a fixed ratio of volume truncation. 
For instance, if one throws away the walker with the highest 
energy in each iteration, then at the $i$-th iteration, nested sampling 
yields a volume $V(E_i) = (1/N)^i \cdot V_0$, where the energy $E_i$ is
determined by the energy of the walker that is currently thrown away.
In contrast, our algorithm would yield a volume $V(E_i) = \prod (n_i/N) \cdot V_0$.   
Thus, any sampling fluctuation reflects on $E_i$ in nested sampling,
whereas it reflects on $V(E_i)$ in our implementation.

Second, nested sampling does not "relax" the walkers as we did in Step 2.
Rather, it duplicates a walker with energy smaller than $E_i$ randomly 
and propagates the cloned walker until it becomes uncorrelated from the original one. 
This operation seems to be more efficient than ours, but as our 
calculation is truncated at $E^{*}$ rather than proceeding to the lowest potential
energy, the number of iterations required is much smaller.  
Third, when applying nested sampling to integrate over the Boltzmann factor,
more walkers are needed in order to better resolve the density of state in energy.
Since PFE does not require any such integration, we can use much less walkers or even
more aggressive energy truncation to calculate $V(E^{*})$.

\section{Results and Discussion}

\subsection{Harmonic Oscillator}
\begin{figure}
\includegraphics[width=0.5\textwidth]{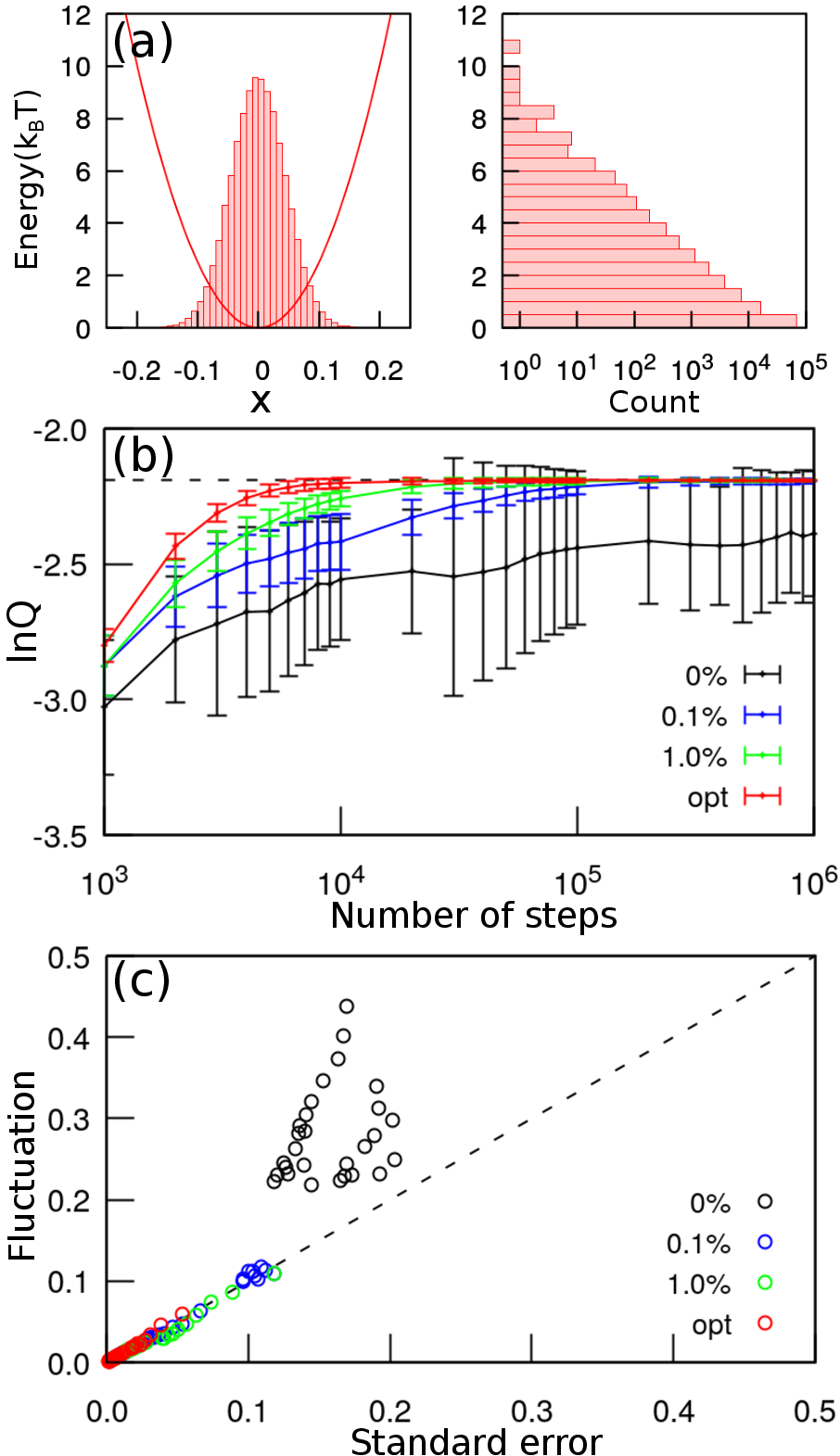}
\caption{\label{fig:1DHO}
(a) The potential energy curve of a harmonic oscillator and the corresponding spatial and energy 
distributions. As anticipated with finite sampling, the spatial probability distribution follows
a Gaussian function (left figure), while the histogram diminishes as the energy level rises 
(right figure), demonstrating that the sampling in the higher energy range is insufficient.
(b) $\ln Q$ computed using PFE (Eq.~\ref{eq:pfe}). Shown are results obtained using different $E^{*}$,
including those resulting in the removal of $0\%$, $0.1\%$, $1.0\%$ of data, as well as $E^{*}$
determined using Eq.~\ref{eq:E*} (labeled as ``opt").
The dashed line indicates the exact solution.
(c) A comparison between the predicted standard error of $\ln \langle f(E;E^{*}) \rangle$ and 
the fluctuation (standard deviation) of $\ln Q$ observed across 100 independent samplings.
A linear relationship is observed in nearly all datasets, except for the black ones.}
\end{figure}

Our first example is the one-dimensional harmonic oscillator potential,
$U(x)=kx^2/2$.  Here we set the force constant as $k=300$ and $k_BT=0.59616$. 
The potential energy curve and the corresponding distributions are depicted in Fig.~\ref{fig:1DHO}(a).  
In this system, we employ a simple Metropolis Monte Carlo algorithm for sampling, utilizing a total
of $10^6$ steps and a step size of $0.1$. 
The trajectory and sampled potential energy are recorded every 10 steps. 
Hundreds of independent samplings are conducted and processed using PFE (Eq.~\ref{eq:pfe}) 
to derive $\ln Q$. The corresponding average and standard deviation (fluctuation) are then depicted 
in Fig.~\ref{fig:1DHO}(b). 
Here, different $E^{*}$ are selected to demonstrate their impact on PFE. The term $a\%$ indicates 
the criterion for selecting $E^{*}$: it is chosen such that for the top $a\%$ of samples in energy,
the corresponding value of $f(E;E^{*})$ is zero.

The outcomes obtained from the optimal $E^{*}$ (calculated using Eq.~\ref{eq:E*}) are labeled as 
``opt" (red curve). As expected, utilizing PFE (Eq.~\ref{eq:pfe}) using the best choice of $E^{*}$
(Eq.~\ref{eq:E*}) leads to a converged result after approximately $10^4$ steps, while the sampling
fluctuation diminishing rapidly with the step count. 
Conversely, the $\ln Q$ calculated using the maximum energy sampled as $E^{*}$ (black curve, $0\%$) 
exhibits substantial fluctuation and fails to converge.  However, upon applying the cutoff 
through the Heaviside function, PFE results converge towards the exact solution (dashed line), 
albeit with slightly inferior performance compared to using the optimal $E^{*}$.
This behavior aligns with the notion that high-energy microstates are often undersampled,
leading to convergence challenges.

The standard error of $\ln \langle f(E;E^{*}) \rangle$ can also be calculated as shown in 
Eq.~\ref{eq:variation} and compared with the fluctuation determined from the standard deviation of 
$\ln Q$ across the 100 sampling repetitions.
These results are displayed in Fig.~\ref{fig:1DHO}(c).
With the exception of the $0\%$ cutoff, the predicted standard error closely matches the fluctuation,
and increasing the sample size results in diminished error. 
This intriguing finding indicates that the fluctuation primarily arises from
inaccuracies in computing $\ln \langle f(E;E^{*}) \rangle$, as the $\ln V(E^{*})$ can be fairly
accurately computed from the trajectory histogram using 100 bins.

\subsection{Double Well Potential}
Next we consider a double well potential defined as follows,
\begin{eqnarray}
\label{eq:dw}
U(x) = \frac{16h}{x_0^4} x^2 (x-x_0)^2 \;, \nonumber
\end{eqnarray}
where $h$ is the barrier height and $x_0$ is the position of the second minimum.
In this example, $k_BT$ is maintained at $0.59616$, while $h$ and $x_0$ are configured to $10k_BT$ 
and $3$, respectively, to introduce some sampling challenges.  For this system, a total of $10^6$ 
steps are sampled using both Monte Carlo and replica-exchange (RE)~\cite{Sugita1999} techniques.  
The latter is a typical enhanced sampling method that simulates multiple replicas with
various temperatures simultaneously. The coordinates between different replicas can be
swapped to enhance the sampling efficiency. 
In the current calculation, 10 replicas are employed in total. The temperatures are linearly
scaled, with the lowest and highest $k_BT$ values set to $0.59616$ and $1.9872$, respectively. 

\begin{figure}
\includegraphics[width=0.5\textwidth]{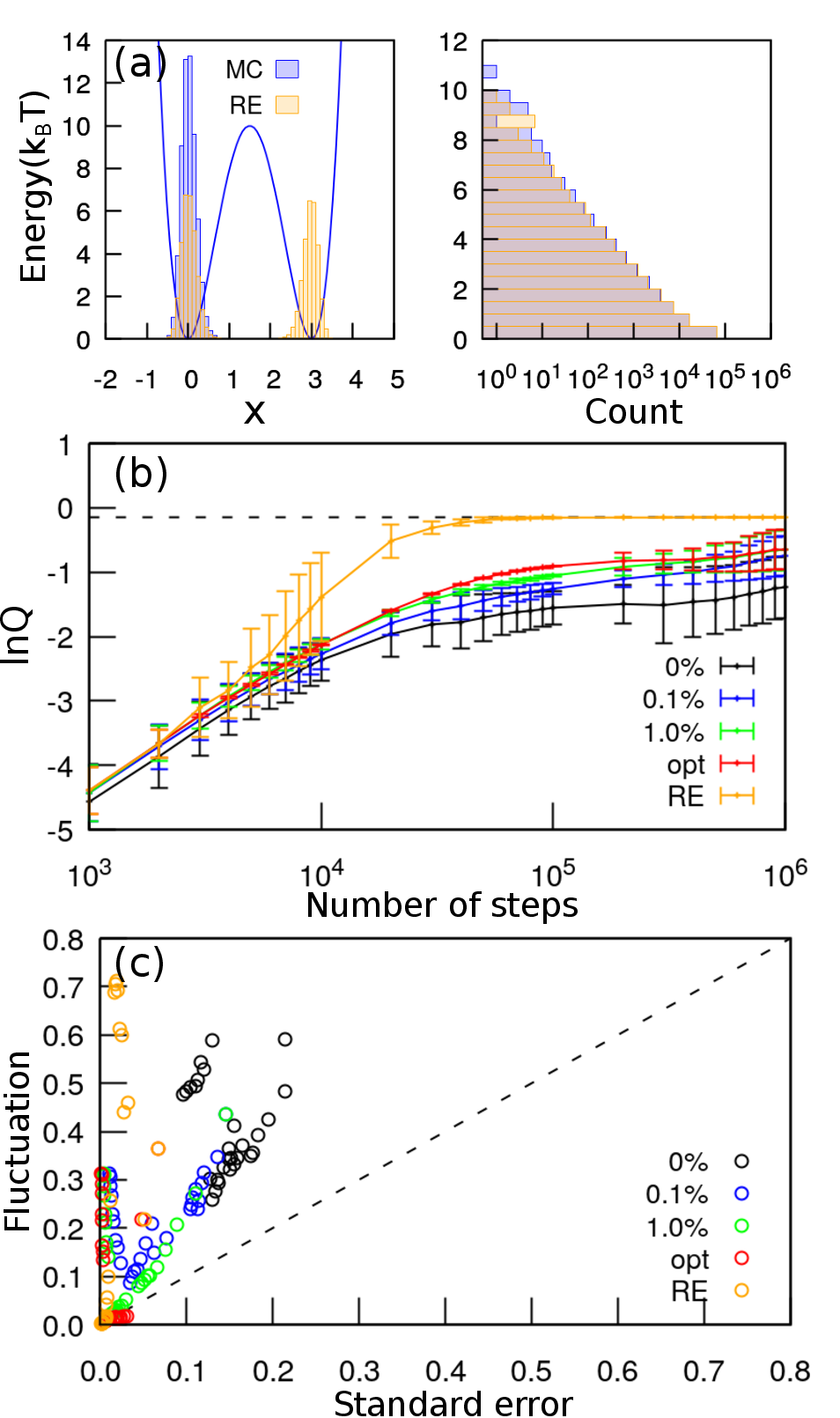}
\caption{\label{fig:1DDW}
(a) The double-well potential energy curve and the corresponding spatial distributions (left) 
and energy histograms (right). Results from a basic Monte Carlo sampling are represented in blue,
while those from replica-exchange (RE) sampling are shown in orange.
(b) $\ln Q$ for the double-well potential, computed using PFE and Monte Carlo sampling. 
Reported are the average and standard deviation from 100 independent calculations.
Labels such as $0\%$, $0.1\%$, $1.0\%$ denote the percentages of data removal, 
employed to determine $E^{*}$.  
The outcome derived from RE sampling and the optimal $E^{*}$ is also depicted (orange).
The numerical exact value (-0.15) is indicated by the dashed line.
(c) A comparison is made between the anticipated standard error of $\ln \langle f(E;E^{*}) \rangle$ and 
the fluctuation of $\ln Q$ across 100 independent samplings.}
\end{figure}

The potential curve and the corresponding distributions are depicted in Fig.~\ref{fig:1DDW}(a).
Due to the substantial barrier, Monte Carlo sampling encounters difficulties in traversing the barrier,
as evident in the blue distribution. In contrast, RE sampling facilitates frequent barrier crossings, 
resulting in a symmetric spatial distribution even for the replica with the lowest temperature, 
as observed in the orange distribution. The energy histograms exhibit similarities between both
sampling methods, except in the high-energy region.
Shown in Fig.~\ref{fig:1DDW}(b) are $\ln Q$ calculated using PFE. 
With the exception of the curve labeled ``RE", all results are obtained from an extensive
Monte Carlo simulation. Once again, the optimal $E^{*}$ (opt) provides the most accurate estimation
of $\ln Q$, although the result is not yet converged after $10^{6}$ sampling steps. 
Owing to the sampling difficulty posed by the high barrier, the $\ln Q$ plot displays a plateau 
around $10^5$ steps ($\ln Q=-0.9$), as depicted by the red curve. 
This plateau arises because the simulation struggles to surmount the barrier (Boltzmann factor 
$e^{-10}= 4.54*10^{-5}$) in shorter sampling durations. 
However, with longer sampling durations, simulations eventually overcome the barrier,
exploring the second well. Consequently, the fluctuation increases, and $\ln Q$ approaches 
the exact solution.

In contrast, $\ln Q$ calculated with RE sampling and the optimal $E^{*}$ shows an early 
convergence well before $10^5$ steps (orange curve).
This observation underscores how sampling difficulties can influence the convergence rate
of calculations. However, as enhanced sampling methods are extensively documented 
in literature~\cite{Sugita1999,Laio2002} and out of scope of this study,  our discussion centers
on the behavior of PFE.
Interestingly, when the sampling remains confined to one well 
(reflected in the plateau-like $\ln Q$ values),the deviation from the exact solution is around $0.7$.
This magnitude is about the same as $\ln(2)$, implying that the primary source of error stems
from halving the size of $V(E^{*})$ when the simulation is trapped in one well.
Introducing enhanced sampling techniques here primarily aids in the accurate computation
of $V(E^{*})$, rather than to $\langle f(E;E^{*}) \rangle$.

In Fig.~\ref{fig:1DHO}(c), a comparison is made between the standard error of 
$\ln \langle f(E;E^{*}) \rangle$ and the fluctuation (standard deviation) of $\ln Q$.  
Unlike the previous example of harmonic oscillator, the fluctuation surpasses the standard error, 
suggesting that calculating $\ln V(E^{*})$ can introduce a significant error, even when directly assessed
from the trajectory.
Nonetheless, increasing the number sampling steps can reduce the fluctuation, as 
demonstrated by the orange curve in panel(b), where the fluctuation eventually dwindles to insignificance.

\subsection{M\"uller-Brown Potential}
\begin{figure}
\includegraphics[width=\textwidth]{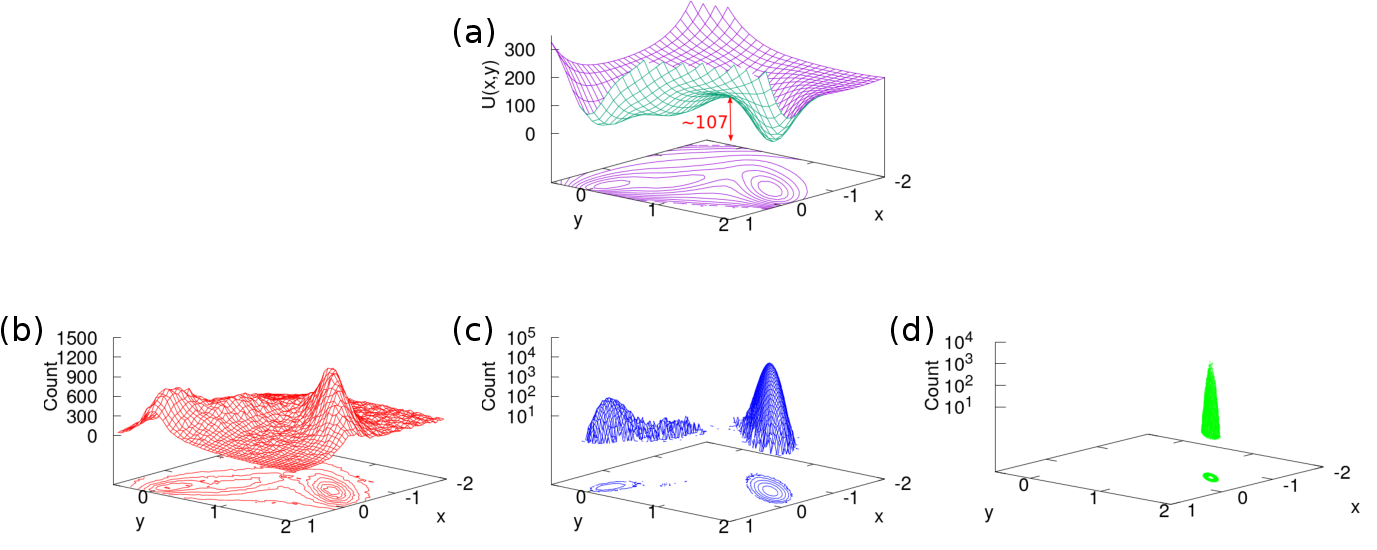}
\caption{\label{fig:MB}
(a) The surface plot of the M\"uller-Brown potential, with a vertical shift to align the 
minimum at 0. The primary barrier height measures around 107, while the secondary barrier height is 
approximately 34.  Both the potential energy and $k_BT$ are represented in the same arbitrary unit. 
(b) Histogram of a single sampled trajectory at $k_BT=100$. The population effectively covers all 
three minima, indicating adequate sampling.
(c) Histogram of a trajectory sampled at $k_BT=10$. The population at the higher energy minima
is non-negligible, but considerably lower than that at the global minimum. Sampling is marginally sufficient in this scenario. 
(d) Histogram of a trajectory sampled at $k_BT=2$. The population is distinctly confined to the deepest well, showcasing entrapment at the low energy region.}
\end{figure}

Another often used model potential is the M\"uller-Brown potential, defined as follows:
\begin{eqnarray}
\label{eq:MB}
U(x,y) = \sum^4_{k=1} A_k \exp \left[ a_k(x-x^0_k)^2+b_k(x-x^0_k)(y-y^0_k)+c_k(y-y^0_k)^2 \right] + U_0 \nonumber \;,
\end{eqnarray}
with $A=(-200,-100,-170,15)$, $a=(-1,-1,-6.5,0.7)$, $b=(0,0,11,0.6)$, $c=(-10,-10,-6.5,0.7)$.
In the original M\"uller-Brown potential, $U_0=0$. Here we set $U_0=147.70$ to ensure  
the potential minimum is at zero. The potential energy surface is shown in Fig.~\ref{fig:MB}(a).
The potential features three local minima, with the highest barrier approximately at 107.
To evaluate the performance of PFE, three different temperatures are employed:
$k_BT=100$, $k_BT=10$, and $k_BT=2$, which correspond to a well-sampled trajectory, 
a poorly sampled trajectory, and a trapped trajectory, respectively. 
The trajectory histograms (proportional to the population) are depicted in Fig.~\ref{fig:MB}(b-d).
It is important to note that in panels (b)-(d), the number of bins employed in each direction is 100, 
but the bin width is automatically adjusted, resulting in varying count scales across the panels.

\begin{figure}
\includegraphics[width=\textwidth]{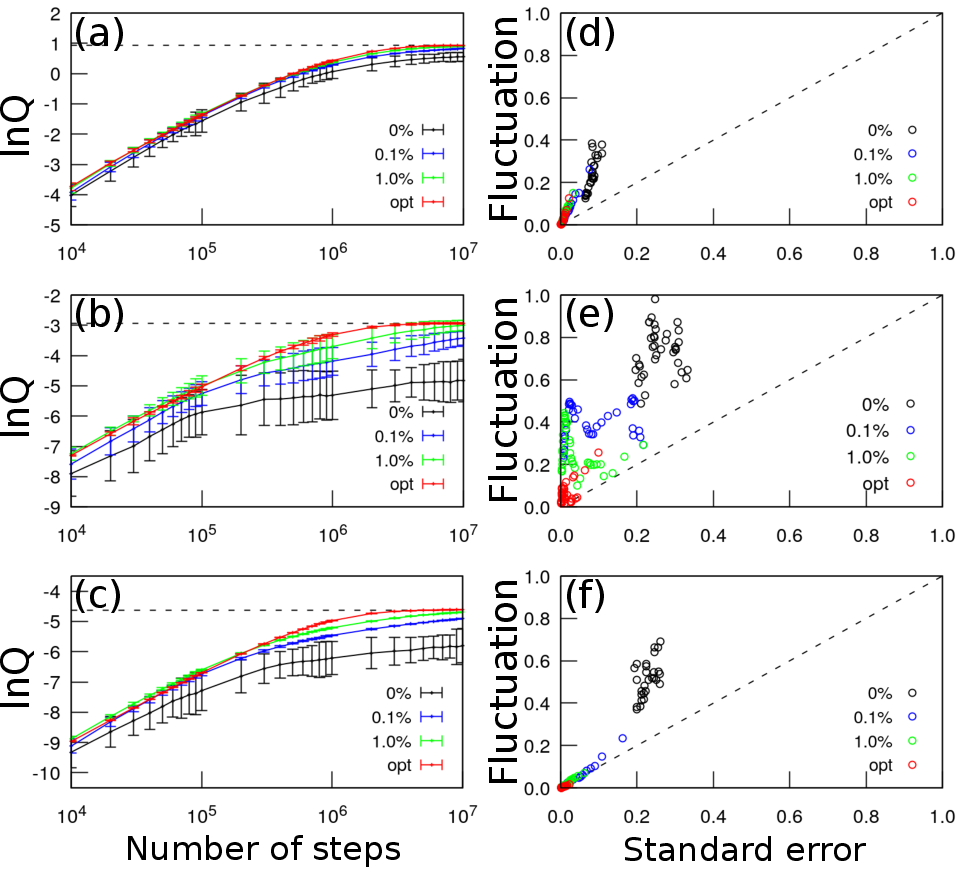}
\caption{\label{fig:2DMB}
(a) - (c): $\ln Q$ for M\"uller-Brown potential, computed using PFE for $k_BT=100$, $k_BT=10$, 
and $k_BT=2$, respectively. The mean and standard deviation from 100 independent calculations are
reported. The dashed line represents the exact solution obtained from a numerical integration. 
PFE with the optimal $E^{*}$ exhibits minimal fluctuation, which is further reduced with an increase
in the number of steps.
(d) - (f): the standard error of $\ln \langle f(E;E^{*}) \rangle$ compared with the fluctuation of $\ln Q$
for $k_BT=100$, $10$, and $2$, respectively. Owing to sampling challenges, the fluctuation is notably
larger than the standard error for $k_BT=10$. See the text for further discussion.}
\end{figure}

The Monte Carlo simulation is once more employed to compute $\ln Q$ (conducted over $10^7$ steps with
data saved every 10 steps). The volume $V(E^{*})$ is determined based on the 2-dimensional 
trajectory histogram, where the area of the populated bins is summed to give $V(E^{*})$. 
Results are illustrated in Fig.~\ref{fig:2DMB}.  Like before, 100 independent trajectories are
processed to derive the mean and standard deviation (fluctuation). Various $E^{*}$ values are
employed to showcase the effect of incorporating the Heaviside function to truncate the inverse 
Boltzmann factor: PFE with the optimal $E^{*}$ (opt) consistently yields the best results, while PFE 
with the highest sampled energy ($0\%$) performs the poorest.
That said, PFE with the optimal $E^{*}$ steadily converges to the exact solution, exhibiting
smaller fluctuation compared to choices of $E^{*}$.
This underscores the significance of selecting $E^{*}$ to minimize error (as per Eq.~\ref{eq:E*}).

Noteworthy observations arise when contrasting the standard error of $\ln \langle f(E;E^{*}) \rangle$
with the fluctuation of $\ln Q$: At low temperature ($2k_BT$, panel f), the sampling is notably 
localized in space, enabling an accurate calculation of $\ln V(E^{*})$.  
Consequently, calculating $\ln V(E^{*})$ scarcely 
contributes to the error in $\ln Q$, with data predominantly aligning along the plot's diagonal. 
As the temperature rises ($10k_BT$, panel e), the sampling ventures beyond barriers to explore other 
local minima. However, the sampling proves insufficient, as indicated by the histogram in Fig.~\ref{fig:MB}(c).
This results in a larger fluctuation, since calculating $V(E^{*})$ based on the 
trajectory can be less precise. Nonetheless, once the convergence is achieved, the fluctuation 
diminishes significantly. 
Finally, at sufficiently elevated temperatures ($k_BT=100$, panel d), the fluctuation and standard 
error once again align, although the alignment is no longer diagonal.
This finding demonstrates that calculating $V(E^{*})$ entails inherent errors, although they are
diminishable with an increasing number of steps, as exemplified by the red curve in panel (a).

\subsection{Lennard-Jones Particles}
The last example to be discussed in a more realistic system: Lennard-Jones particles in a 3-dimensional
box with periodic boundary conditions.  In this example, the box length is fixed at 25 \AA{} and
Lennard-Jones particles are sequentially inserted into the box. 
The parameters are taken from the OpenMM example (Ar atom)~\cite{OpenMM}:
$\epsilon = 0.238$~kcal/mol, $\sigma = 3.4$~\AA{}, with the potential shifted to zero at $3\sigma$ 
for cutoff. The temperature is set to 120~K. The dispersion correction is disabled for simplicity.
Once again we calculate $\ln Q$ using PFE with the optimal $E^{*}$. A basic Monte Carlo simulation 
is employed for data collection. The simulation consists of a $50000$-step equilibration 
with a step size $1.0$ and a $10^6$-step production run, with the data being saved every 1000 steps.
The volume $V(E^{*})$ is determined using the modified nested sampling algorithm (an iterative
Monte Carlo integration method). This approach involves 200 walkers, 2000 equilibration steps, 
and an energy fraction 0.99 at each iteration. The relaxation steps and the number of walkers 
to be relaxed are dynamically adjusted during execution.   
Moreover, the modified nested sampling algorithm is utilized for the direct integration of $\ln Q$,
using identical parameters as those for $V(E^{*})$ calculations. 
Additionally, the standard mFEP-MBAR method~\cite{mFEP,MBAR} is applied for $\ln Q$ calculation
as a reference.
Briefly, mFEP-MBAR computes the free energy difference between two states, e.g. $\ln (Q_N/Q_{N-1})$
with $N$ stands for the number of particles. Since $Q_1$ is simply the box volume, subsequent values
$\ln Q_2$, $\ln Q_3$, ..., $\ln Q_N$ are derived by accumulating the mFEP-MBAR outcomes.
The mFEP-MBAR calculation is conducted using molecular dynamics and alchemical functions implemented 
in OpenMMTools~\cite{OpenMMTools}.  
Simulation parameters are meticulously selected to ensure a comparable level of precision to PFE.
This leads to a maximum number of steps 21210000 (21 windows, each comprising 10000 equilibration steps 
and 1000000 steps production run). 

\begin{figure}
\includegraphics[width=\textwidth]{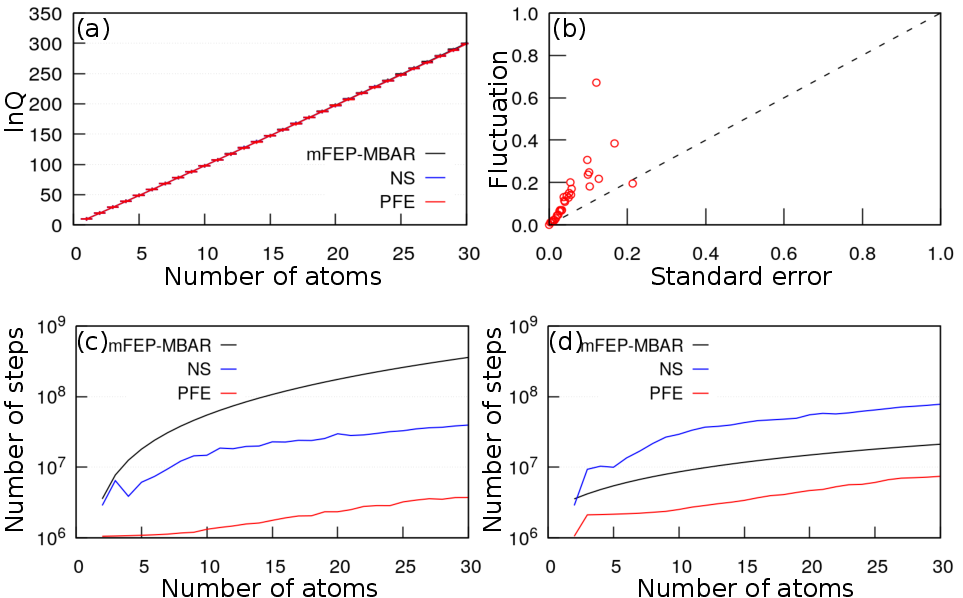}
\caption{\label{fig:3DLJ}
(a) $\ln Q$ calculated using different methods. The standard mFEP-MBAR served as the reference,
and NS denotes data obtained from a numerical integration via the modified nested sampling algorithm.
Reported are the averages and the standard deviations from 10 independent calculations. It is evident 
that the standard deviation is negligible, suggesting a single calculation is sufficient to obtain an accurate $\ln Q$.
(b) The fluctuation and standard error of PFE. (c) The average number of steps required to calculate 
$\ln Q$ for each method. As mFEP-MBAR evaluates $\ln (Q_N/Q_{N-1})$, its $\ln Q$ computation is achieved cumulatively. (d) The average number of steps required to calculate $\ln (Q_N/Q_{N-1})$, viz. the free
energy difference between two states. This is the typical operation for which mFEP-MBAR is designed, 
although surprisingly PFE does not perform worse.}
\end{figure}

Results of $\ln Q$ are depicted in Fig.~\ref{fig:3DLJ}(a). This time, 10 independent calculations are
carried out for each method to determine the mean and the corresponding standard deviation (fluctuation). 
All three methods yield identical $\ln Q$, which is a straight line that increases with the number of
particles. This outcome is anticipated since the slope provides the chemical potential of the system.  
It is noteworthy that the standard deviation is significantly smaller compared to $\ln Q$, implying that
a single calculation is adequate to obtain a precise $\ln Q$. 
The multiple independent calculations are primarily conducted to assess the magnitude of $\ln Q$ 
fluctuations, which are slightly larger than the anticipated standard error, as depicted in 
Fig.~\ref{fig:3DLJ}(b).
Considering performance metrics, a comparison is made based on the number of steps required to 
complete the calculations.  Given the use of distinct sampling techniques, viz. Monte Carlo for
PFE and molecular dynamics for mFEP-MBAR, the average number of sampling steps over 10 computations 
is reported.
Fig.~\ref{fig:3DLJ}(c) presents the step count necessary for $\ln Q$ calculation.
While a direct numerical integration over the Boltzmann factor (labeled NS, blue) is expected
to be computationally expensive, it seems odd that mFEP-MBAR demands even more steps.  
This discrepancy arises because mFEP-MBAR is designed to efficiently compute the free energy difference
between two states, such as $\ln (Q_N/Q_{N-1})$. To get $\ln Q$, 
the free energy differences are accumulated
over $N$, resulting in a higher total step count.
What should really surprise the reader is when one compares the step counts needed to compute
$\ln (Q_N/Q_{N-1})$, i.e. the free energy difference between two states. 
In this instance, PFE's performance slightly surpasses that of mFEP-MBAR, as evident from
the red and black curves in Fig.~\ref{fig:3DLJ}(d).
This intriguing outcome highlights that calculating $\ln Q$ at a given temperature can really be achieved
with the same computational effort as finite sampling.

\begin{figure}
\includegraphics[width=0.5\textwidth]{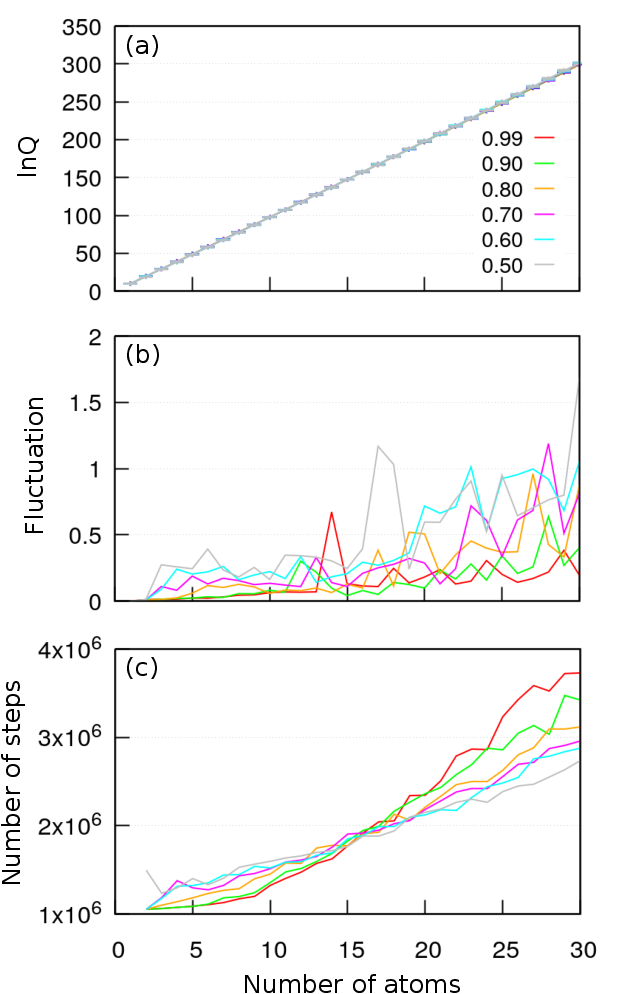}
\caption{\label{fig:3DLJ_PFE}
(a) $\ln Q$ calculated using different fractions for computing $V(E^{*})$. 
The $\ln Q$ values appear to demonstrate stability regardless of the parameter choice.
(b) Sampling fluctuation observed across 10 independent calculations. Employing a more aggressive energy
reduction (e.g. 0.5) for calculating $V(E^{*})$ introduces a larger fluctuation, 
although it remains negligible compared to the value of $\ln Q$. 
The color code is the same as in panel (a).
(c) The computational effort required to calculate $\ln Q$ using different parameters.
The color scheme aligns with that in panel (a).}  
\end{figure}

Last but not least, we discuss the stability of PFE as an estimator.
As shown in Fig.~\ref{fig:3DLJ_PFE}(a), $\ln Q$ is calculated using PFE with varying 
fractions of energy for evaluating $V(E^{*})$ in each iteration. All other parameters are kept 
as previously described. Despite the averages of $\ln Q$ being nearly identical across different 
parameters in 10 calculations, the fluctuation of $\ln Q$ actually increases, see panel (b). 
Additionally, Fig.~\ref{fig:3DLJ_PFE}(c) indicates that adjusting the parameter can marginally
enhance the performance of PFE, although the reduction in effort is not much.

\section{Conclusion}
In this article we proposed a partition function estimator that is composed of an
inverse Boltzmann factor and a Heaviside step function, with a parameter $E^{*}$
that can be determined from minimizing the square of the sampling error.
This results in a working equation that evaluates the expectation value of the proposed
function, along with a volume term $V(E^{*})$ to account for the energy cutoff
imposed by the Heaviside function. We demonstrate that the volume term can be directly 
evaluated from the trajectory for 1- and 2- dimensional examples, while 
a modified nested sampling integration method is proposed for examples with higher dimensions.

As a proof of concept, the performance of the estimator is tested using model examples, 
including the harmonic oscillator potential, the double-well potential, the M\"uller-Brown potential,
and Lennard-Jones particles.  Good agreement with the numerically exact solution is found across
all the test cases. While this result is anticipated, in the case of Lennard-Jones particles, 
a performance comparable to, or even better than, the standard FEP method is noted.

Currently, PFE is not yet ready to handle large biological systems like the standard mFEP-MBAR
can, nor can it calculate the partition function at multiple temperatures in one single run
as nested sampling can. Yet, PFE remains intriguing and warrants further development as it
offers a fresh perspective on addressing a long-standing challenging task.

\begin{acknowledgments}
This work is supported by the Kobilka Institute of Innovative Drug Discovery (KIIDD), School of Medicine, Chinese University of Hong Kong (Shenzhen), and the Shenzhen Science, Technology, and Innovation Commission (SZSTI), grant number JCYJ20230807114206014. 
YCC acknowledges the Royal Society for their support of the precursor to this work (NF171278). Special thanks to Prof. Guanglian Li and Prof. Ye Mei for their valuable discussions, to Prof. Livia B. P\'artay for a fruitful discussion on nested sampling, as well as to Dr. Lantian Yao, Prof. Tzong-Yi Lee, and staff at KIIDD for providing a supportive working environment that nurtured this work.
\end{acknowledgments}

\nocite{*}
\bibliography{PFE}

\end{document}